# The student evaluation of teaching and the competence of students as evaluators[1]

P. Dorta-González [a*] and M.I. Dorta-González [b]

[a] Departamento de Métodos Cuantitativos en Economía y Gestión, Universidad de Las Palmas de Gran Canaria, Gran Canaria, España; [b] Departamento de Estadística, Investigación Operativa y Computación, Universidad de La Laguna, Tenerife, España.

[1] This research has been supported by the Ministry of Science and Technology of Spain under the research project ECO2008-05589.
* Corresponding author. Email: pdorta@dmc.ulpgc.es

When the college student satisfaction survey is considered in the promotion and recognition of instructors, a usual complaint is related to the impact that biased ratings have on the arithmetic mean (used as a measure of teaching effectiveness). This is especially significant when the number of students responding to the survey is small. In this work a new methodology, considering student to student perceptions, is presented. Two different estimators of student rating credibility, based on centrality properties of the student social network, are proposed. This method is established on the idea that in the case of on-site higher education, students often know which others are competent in rating the teaching and learning process.

Keywords: *centrality in social networks, higher education; satisfaction survey, student evaluation of teaching, teaching effectiveness.*

# 1. Introduction

Education has great influence in the economic and social development of countries. Governments set aside resources for improving educational system results, expecting that these contribute to an increase in national wealth and social welfare. Since resources are limited, an adequate educational policy based on quality criteria is required.

Over the last few decades, the quality of teaching and learning in colleges and universities has become an issue of growing concern in many countries around the world (Chen & Hoshower, 2003; Slate et al., 2011). The evaluation of teaching effectiveness is a highly complex process, given that this concept is both subjective and multidimensional. In this sense, many higher education institutions have established quality management systems and are making continuous efforts to ensure and improve the quality of teaching and learning. For this reason the *Student Evaluation of Teaching* (SET) has been extensively researched (Algozzine et al., 2004; Clayson, 2009; Wachtel, 1998).

There are many ways of evaluating educational activity and thus the teaching staff (Berk, 2005). However, obtaining feedback from students is an essential requirement of reflective teaching, allowing teachers to refine their practice and to continue developing as professionals. Many methods can be used to obtain feedback, but the literature suggests that satisfaction surveys predominate (Frick et al., 2009; Kember & Leung, 2009) and student ratings are used as one, sometimes the only and often the most influential, measure of teaching effectiveness (Harvey, 2003; Kwan, 1999).

Considering the European framework, SET represents a subject of great relevance in the creation of the *European Higher Education Area* (EHEA). In this sense, data from a questionnaire by the European University Association addressed to universities taking part in the Bologna Process, showed that many institutions have taken the opportunity of introducing new quality assurance systems and specific internal evaluation procedures (Crosier et al., 2007).

Although the main objective of the SET is the improvement of the learning process, it is also used in the promotion and recognition of teachers (Denson et al., 2010; Onwuegbuzie et al., 2007). SET provides information to three main groups (Penny, 2003): (a) teachers, who can use the information to improve their teaching; (b) managers, who can use the information for accountability and in promotion and tenure decisions; (c) students, who can use the information when choosing modules and courses. Therefore, an instrument for collecting feedback would meet the needs of all these audiences.

No general consensus has been reached about the validity of the SET (Clayson, 2009; Kogan et al., 2010). Implicit in the literature is the assumption that students answer these anonymous satisfaction surveys honestly.

With respect to the anonymous character of theses satisfaction surveys, making a change from anonymous to confidential has been suggested in order to investigate particularly high or low ratings. As Kogan et al. (2010) and Wright (2006) pointed out, when anonymous, students take no responsibility for their ratings. It also eliminates the possibility of follow-up on the results. Therefore, there is no way of determining if students who gave poor ratings were present for most of the class periods or were performing well in class.

In relation to the honesty of students answering theses satisfaction surveys, relatively few studies have attempted to ask the students their general attitudes toward the evaluation, how conscientiously they respond to the questions, and how seriously they take the whole process (Spencer & Schmelkin, 2002). Even fewer studies have attempted to analyse the factors that influence student attitudes towards teaching evaluations, or have examined the behavioural intention of students participating in the evaluation (Chen & Hoshower, 2003). In this sense, students have indicated they are sceptical about the use of satisfaction surveys and consequently do not pay much attention to the ratings (Spencer & Schmelkin, 2002). Therefore, there is a need to work on student attitudes, to motivate and to convince them that their opinions do matter (El Hassan, 2009).

On the other hand, some instructors express mistrust at being evaluated by students (Penny, 2003). This is especially relevant because the effectiveness of the evaluation process depends on a large measure, on the degree of teacher involvement. The mistrust comes, at least in part, from the variability in the students' rating. In this sense, as the empirical application in section 5 shows, the effect of biased ratings over the arithmetic mean is very significant. Therefore, the identification and correction of biased ratings is a relevant and open problem. In this context, it is also reasonable to use the satisfaction surveys to estimate the competence of students as evaluators, and consider this information as a weight in the evaluation process.

In this work, a new methodology considering student to student perceptions is presented. This confidential procedure considers the opinion of the students in order to check the attitude and honesty of students answering the satisfaction survey. Moreover, this method tries to detect and minimize the possible presence of biased ratings in order to increase the teacher involvement. As alternative to the arithmetic mean, in this work two different weighted means, based on centrality properties of student to student competence perceptions, are proposed.

## 2. Centrality measures in social networks

A social network is a structure formed by people (actors) and their relationships (ties). It is represented by a graph with nodes (or vertices) and connections between pairs of nodes, called arcs. Within graph theory and network analysis, there are some measures of centrality that determine the relative influence of a node in the graph.

*Degree centrality* is defined as the number of ties inciding upon a node (rating). If the network is directed, then there are two measures of degree centrality, namely indegree and outdegree. Indegree is the number of arcs directed to the node, and outdegree is the number of arcs that the node directs to others.

*Eigenfactor centrality* assigns relative ratings to all nodes in the network based on the principle that connections to high-rating nodes contribute more to the rating of the node in question than equal connections to low-rating nodes.

While degree centrality assumes that all nodes in the network have the same weight, the eigenfactor centrality gives each node a weight that is proportional to the weights of the adjacent nodes. Note that this definition is recursive.

Centrality measures in social networks have been used in different contexts. The Eigenfactor Metric (Bergstrom, 2007) is a measure of journal influence, recently introduced in the *Journal Citation Reports*. Unlike traditional metrics, such as the popular *Impact Factor*, the Eigenfactor method weighs citations by the influence of the citing journals. This idea comes from Pinski & Narin (1976) in Bibliometrics, Hubbell (1965) in Sociometry, and Leontief (1941) in Economics. Moreover, Brin & Page (1998) use a similar method to design the popular *PageRank* algorithm in the *Google* search engine. In this algorithm, the relevance of a web page is determined by the number of hyperlinks from other pages, as well as the relevance of the linking pages.

## 3. Degree centrality weighted rating

The aim of the degree centrality method is to estimate the evaluation competence of the students based on student to student direct perceptions. Let $n$ be the number of students responding the satisfaction survey, and $r = (r_1,...,r_n)$ be the *rating vector* received by a teacher (using a Likert scale, for example).

Let $C = (c_{ij})_{i,j=1,...,n}$ be the *student–student competence perception matrix* such that $c_{ij} = 1$, $i \neq j$, when student $i$ assesses student $j$ as competent to rate the teacher. Otherwise, $c_{ij} = 0$, $i \neq j$, indicates either student $i$ assesses student $j$ as noncompetent or student *i* has left this question blank. We omit self-valuations, setting all of the diagonal elements of this matrix to 0 ($c_{ii} = 0$ for all *i*). Therefore, row *i* represents the outgoing valuations of student *i*, and column *j* represents the incoming valuations of student *j*.

Let $s_i = \sum_{j=1}^{n} c_{ij}$ be the sum of the elements of row *i*, i.e. the number of competent students according to *i*. Note that $0 \leq s_i \leq n-1$ since $c_{ii} = 0$. The row sum $s_i$ may vary from one row to another, so we later normalize dividing each row by $s_i$.

A *dangling node* in the competence perception network corresponds to a student that does not assess any other student competent; hence, if *i* is a dangling node, then *i* has not outgoing edges and row *i* has all 0 entries. Then, the matrix *C* is transformed into a *normalized competence perception matrix* $D = (d_{ij})_{i,j=1,\ldots,n}$ such that all rows that are not dangling nodes are normalized by the row sum, that is:

$$d_{ij} = \begin{cases} c_{ij} / s_i & \text{if } s_i \neq 0, \\ 0 & \text{if } s_i = 0. \end{cases}$$

Now, each row in *D* has a sum of 0 or 1, and the total sum of the elements in *D* is $0 < \sum_{i=1}^{n}\sum_{j=1}^{n} d_{ij} \leq n$. Note that $\sum_{i=1}^{n}\sum_{j=1}^{n} d_{ij} = n$ when $s_i \neq 0$ for all *i*. We assume that $\sum_{i=1}^{n}\sum_{j=1}^{n} d_{ij} \neq 0$ because, otherwise, no student is perceived competent by any other and thus this procedure would be irrelevant. As shown below, the sum of a column in *D* is an indicator of the student competence.

We define the *degree centrality weighting factor* of a student $j$ as $w_j = \frac{1}{\sum_{i=1}^{n}\sum_{j=1}^{n} d_{ij}} \sum_{i=1}^{n} d_{ij}$. Then, given a rating vector $r = (r_j)_{j=1,\ldots,n}$, the resulting *weighted rating* is $R_d(r) = \sum_{j=1}^{n} w_j r_j$.

It is easy to prove that the *weighting vector* $w = (w_j)_{j=1,\ldots,n}$ verifies: *(i)* $w_j \geq 0, j = 1,\ldots,n$; *(ii)* $\sum_{j=1}^{n} w_j = 1$. The proof of these properties is direct, because $w_j$ is sum and quotient of nonnegative values, and

$$\sum_{j=1}^{n} w_j = \frac{1}{\sum_{i=1}^{n}\sum_{j=1}^{n} d_{ij}} \sum_{i=1}^{n}\sum_{j=1}^{n} d_{ij} = 1.$$

Note that $w_j$ is bigger when $j$ is declared competent by a large number of students. Moreover, if these students declare competent a small number of others then $w_j$ is bigger too. By contrast, $w_j$ tends to 0 if $j$ is assessed competent by a small number of students which, in turn, declare competent many others. In particular, $w_j = 0$ when $j$ is not assessed competent by any student.

Since $R_d(r)$ is a convex linear combination, the weighted rating has the same scale as $r$. Furthermore, it reduces the contribution of ratings from poorly assessed students.

## 4. Eigenfactor centrality weighted rating

The aim of the eigenfactor centrality method is to estimate the competence of students based on student to student cross perceptions. For this purpose, the matrix $D$ is transformed into the *stochastic competence perception matrix H* in which all rows corresponding to dangling nodes are replaced with the vector of all elements $1/n$. Therefore *H* is row-stochastic, that is, all rows are non-negative and add up to 1.

Following PageRank and Eigenfactor approaches, we consider the *transition matrix P*, a column-stochastic matrix defined as follows:

$$P = \alpha \cdot H^t + (1-\alpha) \cdot T,$$

where *T*, known as *teleportation matrix*, is the order *n* square matrix of all elements $1/n$, and *α* is a parameter usually set to 0.85.

Let *x* be the left eigenvector of *P* associated with the unity eigenvalue, that is, the non-zero row vector *x* such that $x = x \cdot P$. Since the matrix *P* is a primitive stochastic matrix, then by virtue of Perron's theorem for primitive matrices, there exists a unique vector *x*, the *influence vector*, such that *(i)* $x > 0$; *(ii)* $\sum_{i=1}^{n} x_i = 1$; and *(iii)* $x = x \cdot P$ (Pillai et al., 2005). The influence vector corresponds to the left eigenvector associated to the largest eigenvalue of *P*, which is 1, since *P* is a stochastic matrix. Furthermore, the influence vector also corresponds to the fixed point of the linear transformation associated with *P*.

Alternatively, the matrix *P* can be interpreted as the transition matrix of a Markov chain. Since *P* is primitive, the influence vector *x* corresponds to the unique stationary distribution of the Markov chain.

The influence vector *x* contains the factors used to weigh the matrix *D*. Therefore, the normalized eigenfactor of a student *j* is

$$v_j = \frac{1}{\sum_{i=1}^{n}\sum_{j=1}^{n} x_i d_{ij}} \sum_{i=1}^{n} x_i d_{ij} \quad (1)$$

We define the *eigenfactor centrality weighting factor* of a student *j* as (1). Then, given a rating vector $r = (r_j)_{j=1,...,n}$, the resulting *weighted rating* is $R_e(r) = \sum_{j=1}^{n} v_j r_j$, where $x = (x_j)_{j=1,...,n}$ is the normalized left eigenvector of *P* associated to the unity eigenvalue.

It is easy to prove that the *weighting vector* $v = (v_j)_{j=1,...,n}$ verifies: *(i)* $v_j \geq 0, j = 1,...,n$; *(ii)* $\sum_{j=1}^{n} v_j = 1$. The proof of these properties is direct, because $v_j$ is sum and quotient of nonnegative values, and

$$\sum_{j=1}^{n} v_j = \frac{1}{\sum_{i=1}^{n}\sum_{j=1}^{n} x_i d_{ij}} \sum_{i=1}^{n}\sum_{j=1}^{n} x_i d_{ij} = 1.$$

The eigenfactor method uses the structure of the entire student to student competence perception graph. The eigenfactor weight of a student is recursively defined in terms of the weights of the valuing students.

## 5. Empirical application

A growing literature has established that students are frequenting Internet sites in search of information about potential professors (Davison & Price, 2009). Such sites allow college students to anonymously evaluate instructors. RateMyProfessors.com (RMP) is a widely used website (Symbaluk & Howell, 2010) on which students can post their ratings of professors. The RMP site allows for open-ended comments and provides 5-point scales for students to rate the professor's helpfulness, clarity, and easiness, and also report level of interest in the course matter. Overall quality is computed by combining the helpfulness and clarity ratings. RMP and similar websites have generated a great deal of controversy among educators and a debate about the validity of student ratings on RMP. However, the RMP measure of instructor teaching effectiveness is correlated with satisfaction surveys administered by universities (Timmerman, 2008; Coladarci & Kornfield, 2007).

In order to analyze the dispersion in student ratings of professors, we have taken into consideration data by the Department of Mathematics of the University of California, Berkeley. The number of professors and ratings at this department allow us to obtain statistically significant results. Data, we have obtained through the RMP website in January 2012 for instructors with at least 5 ratings, correspond to 2224 ratings of 91 professors.

As indicated previously, the RMP website uses a five-point Likert scale for helpfulness and clarity variables. Therefore, the maximum distance among ratings is 4, and 2 represents half the range of possible values of these variables.

The mode of helpfulness and clarity variables obtained for each professor are shown in Table 1. Notice the *mode* is the value that occurs most frequently in a data set. The number of ratings *n* obtained by each professor varies between 5 and 88. Columns $H_2$ and $C_2$ show the number of ratings with absolute deviation of 2 from the mode of variables helpfulness and clarity, respectively. Columns $H_3$ and $C_3$ show the number of ratings with absolute deviation of 3 or larger from the mode of variables helpfulness and clarity, respectively.

[Table 1 about here]

According to the analysis, 29.18% of ratings are at a distance 2 or larger from the mode of Helpfulness variable, and 26.79% of ratings are at a distance 2 or larger from the mode of Clarity variable. It means that at least one of each four ratings is at a distance from the mode half or larger the range of possible values.

Particularly, in case of the Helpfulness variable, 14.84% of ratings have an absolute deviation of 2 from the mode and 14.34% of ratings have an absolute deviation of 3 or larger from the mode. In the case of the Clarity variable, 13.08% of ratings have an absolute deviation of 2 from the mode and 13.71% of ratings have an absolute deviation of 3 or larger from the mode.

*5.1 An example with n=10*

Suppose a five-point Likert scale, where 5 indicates great satisfaction with the teaching action, and 1 no satisfaction. Let $r = (4,4,3,4,5,4,3,1,5,4)$ be the students' rating vector received by a teacher. Note that $r_8 = 1$ is well below the arithmetic mean 3.7, then this rating is biased. The average without $r_8$ is 4 and therefore the effect of $r_8$ is very important in the arithmetic mean.

We have simulated different scenarios, varying row and column 8 in the student to student competence matrix. In scenarios 1 to 3 some students declare $i$=8 as competent while in scenarios 4 to 6 this student is not declared competent by anyone.

[Table 2 about here]

The results are shown in Table 2. Note that $w_8 = v_8 = 0$ in scenarios 4 to 6 because student #8 is declared noncompetent by everyone. We consider the absolute error as the distance with respect to the mean without the biased rating. Absolute errors of eigenfactor ratings are the smallest in each scenario. Finally, the absolute errors of both centrality ratings are around 90% smaller than the arithmetic mean.

## 6. Conclusions

Quality assurance and continuous evaluation of higher education study programmes is one of the major tasks set for the higher education institutions. The process of academic quality improvement necessarily involves the evaluation of teaching staff, as it is an important element in developing a suitable culture of internal evaluation at universities. In this sense, European standards for quality assurance in higher education, as required by the European Higher Education Area, define the important role of students in the quality assurance process.

The anomalous student ratings identification is not easy to systematize. However, it is possible to reduce the effect of biases in relation to the probability of being an anomalous rating. In this work, two weighting systems based on student perceptions are proposed. Results obtained through this evaluation methodology are a good estimation of those obtained if it were possible to identify and eliminate the anomalous ratings.

Finally, the existence of control mechanisms can also serve as an inhibitor of the type of student behavior that is not aligned with the ultimate purpose of improving teaching quality. This is especially relevant because the arithmetic mean of student ratings is frequently used in the promotion and recognition of teachers.

| | Instructor | n | Helpfulness | | | Clarity | | | | Instructor | n | Helpfulness | | | Clarity | | |
|---|---|---|---|---|---|---|---|---|---|---|---|---|---|---|---|---|---|
| | | | Mode | $H_2$ | $H_3$ | Mode | $C_2$ | $C_3$ | | | | Mode | $H_2$ | $H_3$ | Mode | $C_2$ | $C_3$ |
| 1 | Aganagic, M | 27 | 4 | 2 | 0 | 4 | 6 | 0 | 47 | Manon, C | 8 | 1 | 0 | 3 | 5 | 1 | 3 |
| 2 | Agol, I | 21 | 1 | 0 | 2 | 1 | 5 | 1 | 48 | Metcalfe, J | 5 | 5 | 0 | 0 | 5 | 0 | 0 |
| 3 | Aldi, M | 10 | 5 | 2 | 2 | 4 | 2 | 0 | 49 | Mok, CP | 5 | 5 | 0 | 1 | 5 | 0 | 0 |
| 4 | Auroux, D | 9 | 5 | 0 | 0 | 5 | 0 | 0 | 50 | Neu, J | 52 | 1 | 6 | 16 | 1 | 7 | 18 |
| 5 | Balooch, G | 7 | 2 | 2 | 1 | 4 | 1 | 0 | 51 | Ney, P | 18 | 2 | 2 | 1 | 1 | 3 | 6 |
| 6 | Bergman, G | 19 | 1 | 3 | 2 | 1 | 1 | 2 | 52 | Ogus, A | 35 | 3 | 14 | 0 | 5 | 9 | 14 |
| 7 | Borcherds, R | 59 | 3 | 21 | 0 | 3 | 15 | 0 | 53 | Olsson, M | 38 | 4 | 5 | 1 | 4 | 4 | 0 |
| 8 | Bourgoin, F | 6 | 5 | 0 | 0 | 4 | 0 | 0 | 54 | Pachter, L | 5 | 5 | 1 | 0 | 5 | 1 | 1 |
| 9 | Canez, S | 8 | 5 | 0 | 0 | 5 | 0 | 0 | 55 | Paulin, A | 6 | 4 | 0 | 0 | 3 | 1 | 0 |
| 10 | Carter, E | 5 | 5 | 0 | 1 | 3 | 3 | 0 | 56 | Penneys, D | 5 | 1 | 1 | 2 | 3 | 2 | 0 |
| 11 | Cherkassky, V | 8 | 1 | 2 | 1 | 1 | 1 | 1 | 57 | Persson, PO | 11 | 5 | 1 | 1 | 5 | 2 | 0 |
| 12 | Chorin, A | 24 | 1 | 4 | 6 | 1 | 4 | 3 | 58 | Poonen, B | 28 | 2 | 8 | 7 | 5 | 4 | 10 |
| 13 | Christ, M | 18 | 4 | 1 | 0 | 4 | 0 | 1 | 59 | Pugh, C | 32 | 5 | 9 | 12 | 5 | 4 | 13 |
| 14 | Coleman, R | 6 | 5 | 0 | 3 | 2 | 1 | 0 | 60 | Ratner, M | 88 | 5 | 7 | 16 | 5 | 8 | 14 |
| 15 | Comstock, J | 5 | 3 | 2 | 0 | 4 | 0 | 1 | 61 | Reimann, J | 10 | 5 | 0 | 1 | 5 | 0 | 2 |
| 16 | Daenzer, C | 5 | 5 | 0 | 0 | 5 | 1 | 0 | 62 | Reshetikhin, N | 62 | 4 | 5 | 5 | 4 | 8 | 3 |
| 17 | Diesl, A | 15 | 5 | 0 | 0 | 5 | 0 | 0 | 63 | Rezakhanlou, F | 34 | 5 | 7 | 6 | 5 | 5 | 5 |
| 18 | Evans, LC | 28 | 5 | 2 | 1 | 5 | 1 | 1 | 64 | Ribet, K | 35 | 5 | 5 | 2 | 5 | 7 | 7 |
| 19 | Feldman, F | 13 | 3 | 3 | 0 | 4 | 2 | 1 | 65 | Rieffel, M | 18 | 3 | 6 | 0 | 1 | 3 | 7 |
| 20 | Flenner, J | 5 | 5 | 1 | 0 | 5 | 0 | 1 | 66 | Rycroft, C | 6 | 2 | 2 | 0 | 2 | 1 | 0 |
| 21 | Freedman, D | 10 | 1 | 0 | 0 | 1 | 0 | 0 | 67 | Sarason, D | 29 | 5 | 8 | 5 | 5 | 8 | 7 |
| 22 | Frenkel, E | 40 | 5 | 10 | 6 | 5 | 7 | 3 | 68 | Scanlon, T | 49 | 1 | 6 | 20 | 2 | 11 | 4 |
| 23 | Geba, D | 10 | 4 | 2 | 2 | 1 | 2 | 4 | 69 | Serganova, V | 28 | 5 | 6 | 2 | 5 | 0 | 11 |
| 24 | Givental, A | 64 | 1 | 8 | 9 | 1 | 4 | 5 | 70 | Sethian, J | 39 | 5 | 4 | 5 | 5 | 5 | 8 |
| 25 | Graber, T | 17 | 3 | 2 | 0 | 4 | 4 | 3 | 71 | Sharma, A | 5 | 2 | 1 | 1 | 4 | 1 | 0 |
| 26 | Grunbaum, A | 18 | 3 | 5 | 0 | 3 | 4 | 0 | 72 | Silver, J | 36 | 1 | 8 | 5 | 2 | 2 | 5 |
| 27 | Gu, M | 44 | 1 | 10 | 7 | 2 | 4 | 1 | 73 | Slaman, T | 7 | 5 | 2 | 2 | 3 | 1 | 0 |
| 28 | Gurevich, S | 8 | 1 | 0 | 2 | 1 | 1 | 1 | 74 | Spivak, D | 5 | 5 | 0 | 1 | 5 | 0 | 1 |
| 29 | Haiman, M | 50 | 5 | 5 | 1 | 5 | 1 | 2 | 75 | Stankova, Z | 60 | 5 | 4 | 2 | 5 | 5 | 0 |
| 30 | Hald, O | 64 | 5 | 6 | 9 | 4 | 5 | 4 | 76 | Steel, J | 25 | 1 | 3 | 8 | 2 | 4 | 3 |
| 31 | Harrington, L | 34 | 1 | 3 | 10 | 1 | 6 | 5 | 77 | Strain, J | 10 | 5 | 1 | 0 | 4 | 1 | 0 |
| 32 | Harrison, J | 47 | 1 | 3 | 18 | 1 | 2 | 15 | 78 | Sturmfels, B | 18 | 5 | 4 | 3 | 5 | 3 | 3 |
| 33 | Holtz, O | 16 | 4 | 1 | 2 | 5 | 2 | 2 | 79 | Tataru, D | 19 | 3 | 3 | 0 | 2 | 2 | 4 |
| 34 | Hutchings, M | 48 | 4 | 2 | 1 | 4 | 2 | 0 | 80 | Teleman, C | 19 | 1 | 2 | 3 | 1 | 3 | 1 |
| 35 | Johnson, B | 9 | 5 | 0 | 2 | 5 | 0 | 2 | 81 | Voiculescu, D | 23 | 1 | 2 | 2 | 1 | 4 | 0 |
| 36 | Johnson, T | 5 | 5 | 0 | 0 | 5 | 0 | 0 | 82 | Vojta, P | 20 | 1 | 2 | 2 | 2 | 1 | 0 |
| 37 | Jones, V | 44 | 1 | 9 | 13 | 1 | 11 | 15 | 83 | Wagoner, JJ | 64 | 1 | 11 | 9 | 1 | 7 | 9 |
| 38 | Judson, Z | 5 | 4 | 0 | 0 | 5 | 1 | 0 | 84 | Weinstein, A | 24 | 3 | 6 | 0 | 2 | 6 | 2 |
| 39 | Kahan, W | 5 | 5 | 1 | 2 | 5 | 1 | 1 | 85 | Weissman, M | 5 | 5 | 0 | 0 | 5 | 0 | 0 |
| 40 | Karp, D | 12 | 5 | 0 | 0 | 5 | 0 | 0 | 86 | Wilkening, J | 32 | 5 | 7 | 5 | 4 | 7 | 1 |
| 41 | Kirby, R | 52 | 1 | 4 | 9 | 1 | 4 | 9 | 87 | Williams, L | 8 | 4 | 3 | 1 | 5 | 0 | 4 |
| 42 | Krueger, J | 5 | 3 | 2 | 0 | 4 | 1 | 0 | 88 | Wodzicki, M | 39 | 1 | 4 | 12 | 1 | 3 | 11 |
| 43 | Lam, TY | 18 | 4 | 2 | 1 | 5 | 5 | 3 | 89 | Woodin, WH | 57 | 1 | 14 | 13 | 1 | 7 | 14 |
| 44 | Lim, LH | 5 | 5 | 0 | 1 | 5 | 0 | 0 | 90 | Wu, HH | 60 | 1 | 4 | 15 | 1 | 5 | 13 |
| 45 | Liu, AK | 43 | 1 | 8 | 15 | 1 | 10 | 9 | 91 | Zworski, M | 58 | 3 | 14 | 0 | 4 | 12 | 5 |
| 46 | Liu, A | 13 | 3 | 4 | 0 | 3 | 3 | 0 | | | | | | | | | |
| | *Sum* | *984* | | *132* | *129* | | *123* | *96* | | *Sum* | *1240* | | *198* | *190* | | *168* | *209* |

Table 1. Number of ratings *n*, mode of overall quality variables, and number of greater deviated ratings with respect to the variable mode.

| Scenario | Competence matrix $c_{ij}$ | Weights $w_j$ | $v_j$ | Ratings $r_j$ | Arithmetic mean (error) | Weighted rating $R_d$ (error) | $R_e$ (error) |
|---|---|---|---|---|---|---|---|
| 1 | 0 1 1 1 1 1 **0** 1 1 | 0.1479 | 0.1473 | 4 | **3.7** (0.3) | **3.9614** (0.0386) | **3.9767** (0.0233) |
|   | 1 0 0 1 1 1 **0** 1 0 | 0.1089 | 0.1124 | 4 | | | |
|   | 1 1 0 1 1 1 **1** 1 1 | 0.0437 | 0.0478 | 3 | | | |
|   | 1 0 0 0 1 1 **0** 1 1 | 0.1063 | 0.1059 | 4 | | | |
|   | 1 0 1 1 0 1 **0** 1 1 | 0.1089 | 0.1082 | 5 | | | |
|   | 1 1 0 0 0 1 **0** 0 0 | 0.1248 | 0.1240 | 4 | | | |
|   | 1 1 0 1 1 1 **0** 1 1 | 0.1301 | 0.1328 | 3 | | | |
|   | **0 0 0 0 0 0 0 0 0** | **0.0282** | **0.0203** | **1** | | | |
|   | 1 1 1 1 1 1 **0** 0 1 | 0.1109 | 0.1100 | 5 | | | |
|   | 1 1 0 1 1 0 **1** 1 0 | 0.0904 | 0.0912 | 4 | | | |
| 2 | 0 1 1 1 1 1 **0** 1 1 | 0.1442 | 0.1462 | 4 | **3.7** (0.3) | **3.9653** (0.0347) | **3.9774** (0.0226) |
|   | 1 0 0 1 1 1 **0** 1 0 | 0.1091 | 0.1124 | 4 | | | |
|   | 1 1 0 1 1 1 **1** 1 1 | 0.0504 | 0.0498 | 3 | | | |
|   | 1 0 0 0 1 1 **0** 1 1 | 0.1067 | 0.1061 | 4 | | | |
|   | 1 0 1 1 0 1 **0** 1 1 | 0.1091 | 0.1083 | 5 | | | |
|   | 1 1 0 0 0 1 **0** 0 0 | 0.1234 | 0.1236 | 4 | | | |
|   | 1 1 0 1 1 1 **0** 1 1 | 0.1282 | 0.1321 | 3 | | | |
|   | **1 1 1 1 1 1 0 1 1** | **0.0254** | **0.0197** | **1** | | | |
|   | 1 1 1 1 1 1 **0** 0 1 | 0.1109 | 0.1100 | 5 | | | |
|   | 1 1 0 1 1 0 **1** 1 0 | 0.0925 | 0.0919 | 4 | | | |
| 3 | 0 1 1 1 1 1 **0** 1 1 | 0.1498 | 0.1480 | 4 | **3.7** (0.3) | **3.9819** (0.0181) | **3.9832** (0.0168) |
|   | 1 0 0 1 1 1 **0** 1 0 | 0.1147 | 0.1143 | 4 | | | |
|   | 1 1 0 1 1 1 **1** 1 1 | 0.0393 | 0.0461 | 3 | | | |
|   | 1 0 0 0 1 1 **0** 1 1 | 0.1123 | 0.1080 | 4 | | | |
|   | 1 0 1 1 0 1 **0** 1 1 | 0.0980 | 0.1045 | 5 | | | |
|   | 1 1 0 0 0 1 **0** 0 0 | 0.1290 | 0.1255 | 4 | | | |
|   | 1 1 0 1 1 1 **0** 1 1 | 0.1171 | 0.1282 | 3 | | | |
|   | **1 1 0 1 0 1 0 0 1 1** | **0.0254** | **0.0196** | **1** | | | |
|   | 1 1 1 1 1 1 **0** 0 1 | 0.1165 | 0.1120 | 5 | | | |
|   | 1 1 0 1 1 0 **1** 1 0 | 0.0980 | 0.0939 | 4 | | | |
| 4 | 0 1 1 1 1 1 **0** 1 1 | 0.1521 | 0.1505 | 4 | **3.7** (0.3) | **4.0529** (0.0529) | **4.0420** (0.0420) |
|   | 1 0 0 1 1 1 **0** 1 0 | 0.1131 | 0.1154 | 4 | | | |
|   | 1 1 0 1 1 1 **0** 1 1 | 0.0437 | 0.0480 | 3 | | | |
|   | 1 0 0 0 1 1 **0** 1 1 | 0.1104 | 0.1088 | 4 | | | |
|   | 1 0 1 1 0 1 **0** 1 1 | 0.1131 | 0.1111 | 5 | | | |
|   | 1 1 0 0 0 1 **0** 0 0 | 0.1290 | 0.1270 | 4 | | | |
|   | 1 1 0 1 1 1 **0** 1 1 | 0.1316 | 0.1341 | 3 | | | |
|   | **0 0 0 0 0 0 0 0 0** | **0.0000** | **0.0000** | **1** | | | |
|   | 1 1 1 1 1 1 **0** 0 1 | 0.1151 | 0.1129 | 5 | | | |
|   | 1 1 0 1 1 0 **0** 1 0 | 0.0919 | 0.0921 | 4 | | | |
| 5 | 0 1 1 1 1 1 **0** 1 1 | 0.1480 | 0.1499 | 4 | **3.7** (0.3) | **4.0476** (0.0476) | **4.0413** (0.0413) |
|   | 1 0 0 1 1 1 **0** 1 0 | 0.1129 | 0.1153 | 4 | | | |
|   | 1 1 0 1 1 1 **0** 1 1 | 0.0504 | 0.0490 | 3 | | | |
|   | 1 0 0 0 1 1 **0** 1 1 | 0.1105 | 0.1089 | 4 | | | |
|   | 1 0 1 1 0 1 **0** 1 1 | 0.1129 | 0.1111 | 5 | | | |
|   | 1 1 0 0 0 1 **0** 0 0 | 0.1272 | 0.1268 | 4 | | | |
|   | 1 1 0 1 1 1 **0** 1 1 | 0.1296 | 0.1337 | 3 | | | |
|   | **1 1 1 1 1 1 0 1 1** | **0.0000** | **0.0000** | **1** | | | |
|   | 1 1 1 1 1 1 **0** 0 1 | 0.1147 | 0.1129 | 5 | | | |
|   | 1 1 0 1 1 0 **0** 1 0 | 0.0938 | 0.0924 | 4 | | | |
| 6 | 0 1 1 1 1 1 **0** 1 1 | 0.1536 | 0.1508 | 4 | **3.7** (0.3) | **4.0643** (0.0643) | **4.0436** (0.0436) |
|   | 1 0 0 1 1 1 **0** 1 0 | 0.1185 | 0.1163 | 4 | | | |
|   | 1 1 0 1 1 1 **0** 1 1 | 0.0393 | 0.0473 | 3 | | | |
|   | 1 0 0 0 1 1 **0** 1 1 | 0.1161 | 0.1095 | 4 | | | |
|   | 1 0 1 1 0 1 **0** 1 1 | 0.1018 | 0.1096 | 5 | | | |
|   | 1 1 0 0 0 1 **0** 0 0 | 0.1327 | 0.1276 | 4 | | | |
|   | 1 1 0 1 1 1 **0** 1 1 | 0.1185 | 0.1323 | 3 | | | |
|   | **1 1 0 1 0 1 0 0 1 1** | **0.0000** | **0.0000** | **1** | | | |
|   | 1 1 1 1 1 1 **0** 0 1 | 0.1202 | 0.1136 | 5 | | | |
|   | 1 1 0 1 1 1 **0** 1 0 | 0.0994 | 0.0929 | 4 | | | |

Table 2. Weights and weighted ratings for different scenarios where $r_8$ is a biased rating (the average without the biased rating is 4).